# Experimental Estimate of Beam Loading and Minimum rf Voltage for Acceleration of High Intensity Beam in the Fermilab Booster

Xi Yang, C. Ankenbrandt

*Fermi National Accelerator Laboratory*

Box 500, Batavia IL 60510

J. Norem

*Argonne National Laboratory*

9700 S. Cass Ave, Argonne, IL, 60439

## Abstract

The difference between the rf voltage seen by the beam and the accelerating voltage required to match the rate of change of the Booster magnetic field is used to estimate the energy loss per beam turn. Because the rf voltage (RFSUM) and the synchronous phase can be experimentally measured, they can be used to calculate the effective accelerating voltage. Also an RFSUM reduction technique has been applied to measure experimentally the RFSUM limit at which the beam loss starts. With information on beam energy loss, the running conditions, especially for the high intensity beam, can be optimized in order to achieve a higher intensity beam from the Fermilab Booster.

## Introduction

The energy loss per beam turn in a Booster cycle can be estimated from the difference between the effective accelerating voltage and the accelerating voltage required by dB/dt. The measured synchronous phase ($\varphi_s$) [1] and RFSUM (V) [2] are used to calculate the effective accelerating voltage ($V \times \mathrm{Sin}(\varphi_s)$). RFSUM is the vector sum of rf voltages at the cavity gaps of all Booster rf stations. Since one can vary RFSUM to values below normal at any time, [3] it is useful to determine experimentally the RFSUM values at which the beam loss starts. Furthermore, how much rf voltage is used to provide the rf bucket area, as well as the longitudinal emittance, can be estimated by comparing the RFSUM limit and the effective accelerating voltage. Finally, the optimal running condition can be achieved by modifying the rf amplitude curve, for example by increasing the RFSUM during the transition crossing period to provide sufficient bucket area for a mismatched bunch developed right after transition, especially at high intensity. It may also be possible to decrease RFSUM in the early part of the cycle to maximize the transmission efficiency and control the longitudinal emittance growth at the same time.

## Experimental results

All data were taken under similar conditions with the extracted beam intensity about $5.1 \times 10^{12}$ protons. The synchronous phase was measured using the synchronous phase detector recently developed for the Fermilab Booster.[1] Results are shown in Fig. 1. The method for determining the lower limit for RFSUM is to reduce it slowly to the value at which the beam loss starts. The purpose of slow reduction is to avoid the complication of the beam recapture.[3] In our experiment, the RFSUM limit was measured at 3.5 ms, 6.5 ms, 9.5 ms, 12.5 ms, 15.5 ms, 18.5 ms, 21.5 ms, 24.5 ms, 27.5 ms, 30.5 ms, and 32 ms separately. The beam was injected at 0 ms. RFSUM and RFSUM limit were obtained by digitizing the results shown in Fig. 2. Afterwards, the effective accelerating voltage is calculated using Eq. 1.

$$V_{accelerating}(t) = V_{RFSUM}(t) \times Sin(\varphi_s(t)) \qquad (1)$$

Here, $V_{RFSUM}(t)$ is the RFSUM at time $t$ in a Booster cycle, and $\varphi_s(t)$ is the synchronous phase at time $t$. The RFSUM in a Booster cycle is shown as the black curve in Fig. 3. Also, the RFSUM limit, the effective accelerating voltage, and the accelerating voltage required by dB/dt are shown as red curve, green curve and blue curve separately in Fig. 3. The early fluctuations in the effective accelerating voltage seem real and repeatable, and are associated with the fluctuations in RFSUM. These fluctuations are not well understood. The difference between the effective accelerating voltage (green curve) and the accelerating voltage required by the magnet ramp (blue curve) gives the information to estimate beam loading, as shown in Fig. 4.

The bucket area of the beam can be calculated from the RFSUM limit and beam loading. Because the RFSUM limit is the minimum rf voltage which is required for providing the bucket area as well as compensating the energy loss per Booster turn. Results are shown in Fig. 5. The beam has a nearly constant bucket area of 0.05 eVs before transition and 0.07 eVs after transition. The longitudinal emittance growth is about 40% after transition, and it is smaller than one expects. The effective accelerating voltage is expected to be higher when the beam intensity goes higher, because beam loading linearly increases with the beam intensity, as shown in Fig. 6 [4], while the accelerating voltage required by dB/dt is independent of the beam intensity.

**Comments**

The RFSUM limit and the effective accelerating voltage are nearly equal near transition crossing, as shown in Fig. 3. This implies that more rf voltage will be needed during transition crossing for Booster running at higher intensities. However, there is too much rf voltage in the early part of the Booster cycle, and this is true even at high intensity. One can expect more rf voltage from the Booster rf stations to be available at transition crossing if the rf accelerating voltage is lowered in the beginning of the cycle. Booster could run at higher intensity either if the rf voltage curve is appropriately modified in the present rf configuration or if more rf stations are installed to provide more rf power near transition.

**Acknowledgement**

Special thanks should be given to James MacLachlan. He spent a lot of time in helping authors understand the basic related to this work.

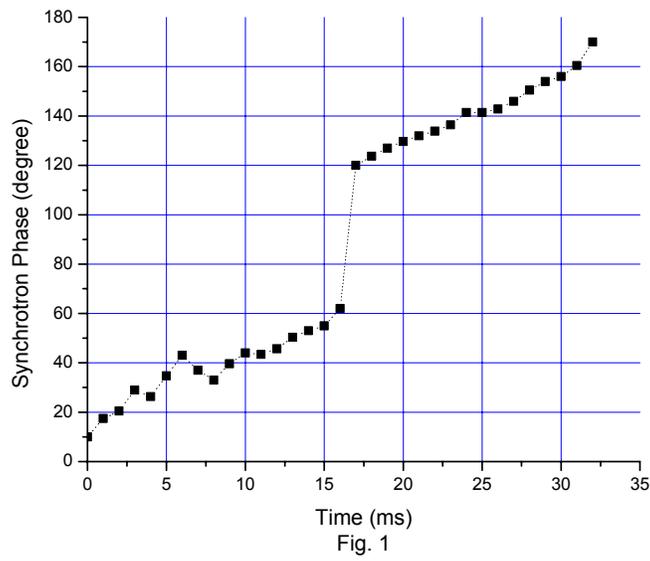

Fig. 1. The synchronous phase measured at the extracted beam intensity of $5.1\times10^{12}$ protons in a Booster cycle.

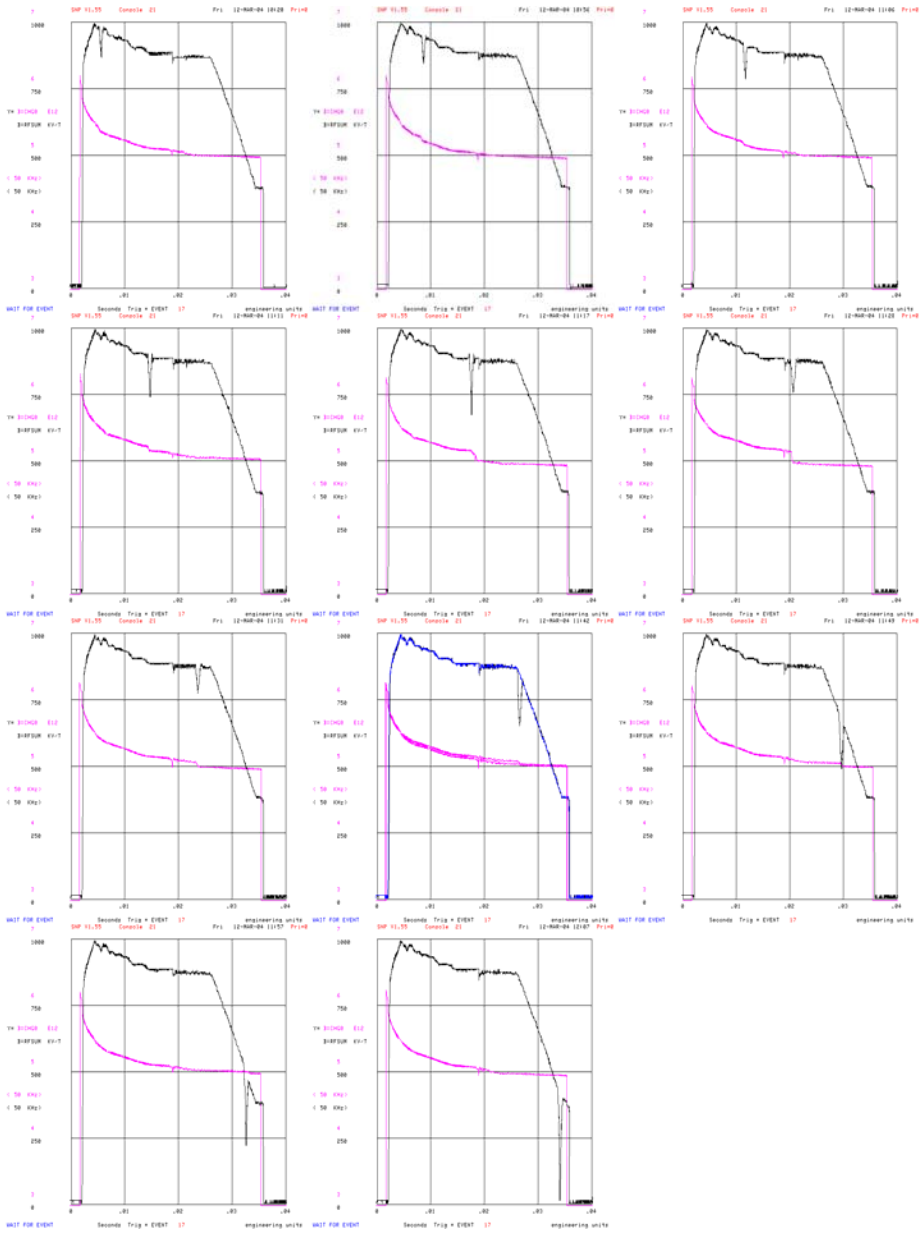

Fig. 2. The RFSUM lower limits measured at 3.5 ms, 6.5 ms, 9.5 ms, 12.5 ms, 15.5 ms, 18.5 ms, 21.5 ms, 24.5 ms, 27.5 ms, 30.5 ms, and 32 ms for the extracted beam intensity of $5.1 \times 10^{12}$ protons. The beam is injected at 0 ms. The black curves represent RFSUM over a Booster cycle, and the magenta curves represent the beam intensity.

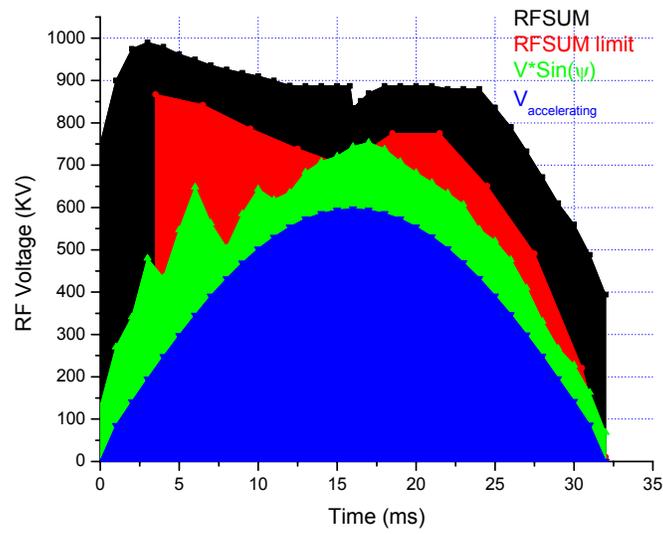

Fig. 3. The black curve represents RFSUM over a Booster cycle, the red curve represents the lower limit for RFSUM, the green curve represents the effective accelerating voltage, and the blue curve represents the accelerating voltage required by the magnet ramp, all data were taken under closely similar conditions at the extracted beam intensity of $5.1 \times 10^{12}$ protons.

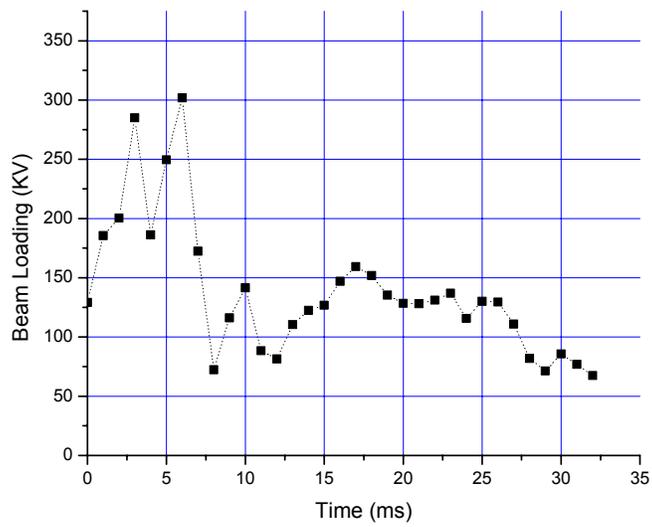

Fig. 4. Experimental estimate of the beam energy loss per turn in a Booster cycle at the extracted beam intensity of $5.1\times10^{12}$ protons.

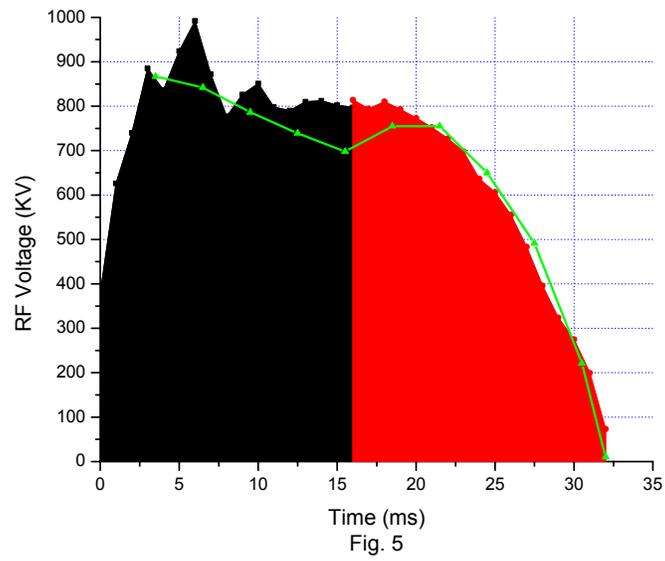

Fig. 5. The black curve represents RFSUM required for a constant bucket area of 0.05 eVs before transition, the red curve represents RFSUM required for a constant bucket area of 0.07 eVs after transition, and the green curve represents the lower limit for RFSUM.

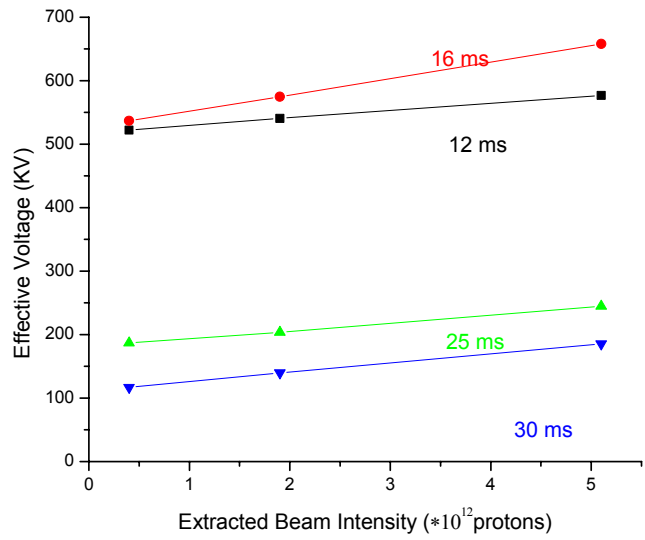

Fig. 6  The effective accelerating voltage *vs.* beam intensity.  The black, red, green, and blue curves were taken at 12 ms, 16 ms, 25 ms, and 30 ms in a Booster cycle.